\documentstyle[twocolumn,pre,aps,epsf]{revtex}

\def\i{{\rm i}}
\def\d{{\rm d}}
\def\e{{\rm e}}
\def\vector#1{{\bf #1}}
\def\vk{{\vector k}}
\def\vq{{\vector q}}

\def\eF{{\epsilon_{\rm F}}}
\def\Tc{{T_{\rm c}}}

\def\hightc{{high-$T_{\rm c}$ }}

\def\nh{{n_{\rm h}}}
\def\nhc{{n_{\rm h}^{\rm c}}}

\def\Tc{{T_{\rm c}}}

\def\Tg{{T_{\rm g}}}

\def\gsim{\stackrel{{\textstyle>}}{\raisebox{-.75ex}{$\sim$}}}
\begin{document}
\draft

\twocolumn[\hsize\textwidth\columnwidth\hsize\csname
@twocolumnfalse\endcsname


\title{
Pseudogap due to Antiferromagnetic Fluctuations and the Phase Diagram 
of High-Temperature Oxide Superconductors}

\author{Hiroshi {Shimahara}, Yasumasa {Hasegawa}$^1$,
Mahito {Kohmoto}$^2$}

\def\runtitle{
Pseudogap due to antiferromagnetic fluctuations
and the phase diagram of \hightc superconductors
}
\def\runauthor{Hiroshi {\sc Shimahara}, Yasumasa {\sc Hasegawa}, 
Mahito {\sc Kohmoto}}

\address{
Department of Quantum Matter Science, ADSM, Hiroshima University,
Higashi-Hiroshima 739-8526, Japan \\
$^1$Faculty of Science, Himeji Institute of Technology,
Kamigouri-cho, Akou-gun, Hyogo 678-1297, Japan \\
$^2$Institute for Solid State Physics, University of Tokyo, 7-22-1
Roppongi, Minato-ku, Tokyo, Japan}

\date{Received ~~~ February 2000}
\maketitle


\begin{abstract}
A reduction of the density of states near the Fermi energy in the normal 
state (pseudogap) of high-temperature oxide superconductors is examined 
on the basis of the two-dimensional tight-binding model with effective 
interactions due to antiferromagnetic fluctuations. 
By using antiferromagnetic correlation lengths which are 
phenomenologically assumed, 
the doping dependence of the pseudogap is obtained. 
The superconducting transition temperature decreases and eventually 
vanishes due to the pseudogap as the hole concentration is reduced. 
\end{abstract}


]

\narrowtext


In high-temperature oxide superconductors (HTSCs), 
behaviors which can be attributed to a reduction of the density of 
states (DOS) near the Fermi energy (pseudogap) have been observed. 
For example, experimental results 
in photoemission spectroscopies~\cite{Mar96,Din96,Har96,Sat99}, 
tunneling spectroscopies~\cite{Ren98,Miy98,Eki99}, 
a specific heat measurement~\cite{Lor93}, 
NMR experiments~\cite{Yas89,War89,Tak91,All93,Ish98}, 
and neutron scattering~\cite{llb}. 
However, the origin of the pseudogap remains controversial: 
preformed Cooper pairs in the spinon condensation 
model~\cite{Suz88,Kot88,Tan94,lee97,Tri95,Eme95}, 
spin fluctuation in the nearly antiferromagnetic (AF)
spin fermion model~\cite{Sch98},
$d$-wave pairing fluctuations~\cite{Eng98,Ici99,Hot99},
AF fluctuation-mediated pairing interactions in the $d$-$p$
model~\cite{Kob99},
$d$-wave pairing fluctuations and AF fluctuations~\cite{Ono99}, 
and so forth.

Among these candidates, the decrease of $\Tg$ (the temperature below which 
the pseudogap phenomena are observed) with hole doping seems to suggest 
a possibility that the pseudogap mainly originates from AF fluctuations 
at least at temperatures much higher than the superconducting transition 
temperature ($\Tc$). 
The AF long-range-ordered phase occurs in the vicinity of the half-filling, 
where the real gap is open. 
It is likely to have a pseudogap structure in the DOS in its proximity due to 
strong AF fluctuations. 
In this mechanism, the decrease of $\Tg$ can be explained naturally, 
since the AF fluctuations decrease with doping. 
On the other hand, if we assume that the pseudogap is due to pairing 
fluctuations, temperature $\Tg$ can be regarded as the temperature 
at which pairing fluctuations begin to occur. 
Therefore, when $\Tg$ increases, it is natural to consider that $\Tc$ should 
increase as well. 
However, such behavior is inconsistent with the observed behavior of the 
opposite doping dependences of $\Tg$ and $\Tc$ in the underdoped region.

In this paper we examine the pseudogap due to AF fluctuations. 
The reduction of the DOS near the Fermi surface should suppress $\Tc$ in 
the underdoped region, and thus $\Tc$ has a peak as a function of hole 
concentration.
We intend to describe a minimum theory which reproduces the phase diagram 
of a HTSC. 
Hence, we omit some of the details which do not change the qualitative 
behavior of $\Tc$. 
For example, we adopt the static approximation for the spin fluctuations. 
The static approximation does not produce the imaginary part of the 
self-energy and the broadening of the one particle weight. 
However, the real part of the self-energy could reproduce 
the reduction of the DOS near the Fermi surface, 
which is the most intuitive definition of the pseudogap. 
Therefore, the static approximation is sufficient for our purpose. 
In addition, Schmalian et al. \cite{Sch98} argued that the 
characteristic frequency of the spin fluctuations $\omega_{\rm sf}$ is 
much smaller than the temperatures of interest for HTSCs.

In our formulation, AF fluctuations are taken into account through 
a renormalization effect. 
The importance of the renormalization effect in the doping dependence of 
the $\Tc$ of HTSC was discussed on the basis of the two-dimensional 
Hubbard model~\cite{Shi88}. 
It was shown that $\Tc$ is reduced considerably near the boundary of 
the AF long-range-ordered phase by the strong renormalization effect 
due to AF fluctuations. 
However, the pseudogap was not taken into account sufficiently.

This work was extended to include the pseudogap due to AF fluctuations 
in a quasi-one-dimensional (Q1D) Hubbard model as a model of Q1D organic 
superconductors~\cite{Shi89}. 
It was shown that the pseudogap suppresses $\Tc$ markedly near the spin 
density wave boundary. 
As a result, phase diagrams of Q1D organic superconductors in the 
pressure and temperature plane were semiquantitatively reproduced.

In HTSCs, however, the same approach based on one of the microscopic 
models is difficult. 
The interlayer coupling is much smaller and the temperature range of 
interest is much higher in HTSCs than it is in the organic 
superconductors. 
Hence, the thermal fluctuations are very strong in HTSCs. 
This situation makes a quantitative argument difficult. 
In addition, there is no common consensus as to which microscopic model 
is appropriate for HTSCs: a single-band Hubbard, $d$-$p$, 
$t$-$J$ models, and so forth. 
Therefore, we treat phenomenologically determined AF fluctuations 
from the experiments instead of calculating them microscopically, 
and concentrate on their qualitative features.

In the calculation of $\Tc$, for simplicity, we assume that the coupling 
constant of the pairing interactions does not depend on the doping. 
Such interactions may be attributed to those mediated by phonons. 
It may appear that the $d$-wave pairing does not occur with 
phonon-mediated interactions. 
However, it is shown that pairing interactions mediated by screened 
phonons can give rise to a $d$-wave pairing superconductivity 
in the presence of AF fluctuations~\cite{Koh99}. 
Here, we do not specify the origin of the pairing interactions. 
To some extent there may be a contribution from the exchange of 
AF fluctuations

We calculate the electron self-energy up to the one-loop approximation as 
\begin{equation} 
     \Sigma_{\sigma}(k) = - T \sum_{n'} N^{-1} \sum_{\vk'} \sum_{\sigma'} 
     V_{\sigma \sigma'}(k,k') G_{\sigma'}(k') , 
     \label{eqSE}
\end{equation}
with $k = (\vk, \i \omega_{n})$, where $V_{\sigma \sigma'}$ is 
the effective interaction due to the exchange of the fluctuations 
and $G_{\sigma}(k')$ is the renormalized electron Green's function, 
\begin{equation} 
     G_{\sigma}(k) 
     = \frac{1}{i \omega_n-\epsilon_{\bf k}-\Sigma_{\sigma}(k)+\mu}.
     \label{eqG}
\end{equation}
We consider a two-dimensional tight-binding model with the electron 
dispersion 
\begin{equation} 
     \epsilon_{\vk} = - 2 t (\cos k_x + \cos k_y) 
                      - 4 t' \cos k_x \cos k_y , 
     \label{eqdispersion}
\end{equation}
where $t$ and $t'$ are the nearest- and next-nearest-neighbor hoppings, 
respectively.

We express the effective interactions due to exchange of magnetic 
fluctuations as 
\begin{eqnarray} 
     V(\vk,\vk') = \frac{V_0}{(\vk-\vk'- \vq_m)^2 + {q_0}^2} , 
     \label{eqVeffective} 
\end{eqnarray}
within the static approximation. 
Here, $V_{0}$ and $q_0$ are the phenomenological parameters. 
Since the effective interactions due to the exchange of magnetic 
fluctuations are proportional to the spin susceptibility $\chi(\vk-\vk')$, 
they must have a sharp peak at $\vk = \vk' + \vq_m$, 
where $\vq_m$'s are the nesting vectors near $(\pi,\pi)$ for the 
AF fluctuations which give the largest value of $\chi(\vq)$.

There are four nesting vectors such as 
$\vq_m = (\pi \pm \delta, \pi)$ or 
$\vq_m = (\pi, \pi \pm \delta)$, 
when the incommensurate AF fluctuations occur~\cite{Has87}. 
In eq.~(\ref{eqVeffective}), 
we take a nesting vector such as $\pi \ge q_{mx} \ge q_{my} > 0$ 
for $k_x-k'_x \ge k_y-k'_y \ge 0$. 
For other regions of $k_x-k'_x$ and $k_y-k'_y$, $\vq_m$ is chosen 
so that $V(\vk,\vk')$ satisfies the symmetry condition. 
The AF correlation length $\xi \equiv 1/q_0$ diverges at the critical 
hole concentration $\nhc$. 
We take 
\begin{equation} 
     \frac{1}{q_0} = \frac{C}{\pi (\mu_{\rm c} - \mu)^{1/2}} , 
     \label{eqAFxi} 
\end{equation} 
which diverges at a critical chemical potential $\mu_{\rm c}$.
For small hole concentrations, the chemical potential $\mu$ is roughly 
proportional to the hole concentration $\nh$. 
Thus, the above form roughly implies $\xi \propto 1/\sqrt{\nh}$, 
which is plausible in the sense that the AF correlation length 
is directly related to the average distance between holes~\cite{Bir89}.

Thus, we finally obtain an equation to solve in a compact form
\begin{equation} 
\renewcommand{\arraystretch}{2.5}
     \Sigma_{\sigma}(\vk)
         = \frac{1}{2} N^{-1} \sum_{\vk'}
             {V(\vk,\vk')
              \tanh(\frac{\epsilon_{\bf k'}
            + \Sigma_{\sigma}({\bf k}')-\mu}{2T})}.
     \label{eqSEtosolve}
\end{equation}
With the static approximation of eq.~(\ref{eqAFxi}), 
we only need to solve the real part of the self-energy.
Thus, it is found that the pseudogap appears as shown below, 
for example, in the DOS, which is given by 
\begin{equation} 
     \rho(\epsilon) = 
       \int \frac{\d k_x \d k_y}{(2\pi)^2} 
       \delta(\epsilon - {\tilde \epsilon}_{\vk}) , 
     \label{eqDOS} 
\end{equation} 
where we have put 
${\tilde \epsilon}_{\vk} = \epsilon_{\vk} + \Sigma_{\sigma}(\vk')$.

We solve the self-consistent equation for large discrete 
$512 \times 512$ momentum points in the first Brillouin zone. 
We confirmed that the results of $512 \times 512$ practically coincide 
with those of $256 \times 256$, within the width of the lines 
in our figures of DOS and $\Tc$. 
Thus, practically, the system size $512 \times 512$ can be regarded as 
being within the thermodynamic limit at the present temperature. 
We interpolate the obtained self-energy linearly in the momentum space 
so that the first Brillouin zone has $4096 \times 4096$ points for the 
calculation of DOS.

We consider the case of nearest-neighbor hoppings 
($t \neq 0$ and $t' = 0$).
We choose $V_0 = -0.4$ and $T = 0.01$, and take $C = 30$ and 
$\mu_{\rm c} = -0.05$ in eq.~(\ref{eqAFxi}). 
We use the units in which $t = 1$. 
The value of $\mu_{\rm c}$ gives $\nhc \approx 0.02 = 2 \%$, where $\nhc$ 
denotes the critical hole concentration of the AF long-range-order.

\begin{figure}[htb]
  \begin{center}
  \leavevmode \epsfxsize=7cm  \epsfbox{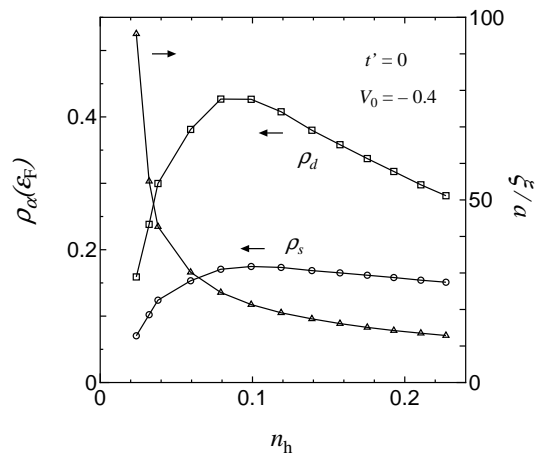}
  \end{center}
\caption{
Doping dependence of the AF correlation length $\xi$ (triangle), 
and the resulting densities of states $\rho_s(\eF)$ and $\rho_d(\eF)$, 
(circles and squares, respectively). 
The solid lines are guides for the eyes. 
$\rho_d(\eF)$ is the effective DOS for the $d$-wave pairing 
which is averaged with the weight $(\cos k_x - \cos k_y)^2$. 
}
\label{fig:rho_and_xi}
\end{figure}

Figure~\ref{fig:rho_and_xi} shows the $\nh$ dependence of the AF 
correlation length $\xi$, and the resulting DOS at the Fermi energy. 
In this figure, $\rho_d$ is the effective DOS for the $d$-wave pairing 
in which the order parameter is 
$\Delta(\vk) \propto (\cos k_x - \cos k_y)$, 
\begin{equation} 
     \rho_d(\mu) =
       \int \frac{\d k_x \d k_y}{(2\pi)^2} \,
       \delta( {\tilde \epsilon}_{\vk} - \mu ) \,
       (\cos k_x-\cos k_y)^2 . 
     \label{eqDOSd}
\end{equation}
$\rho_s$ is that for the $s$-wave pairing, 
($\Delta(\vk) = {\rm const}.$), 
which is equivalent to the usual DOS.

{From} the obtained DOS, we calculate $\Tc$ by the conventional 
weak coupling formula 
\begin{equation} 
     T_{\rm c} = 1.13 \,\, \omega_{\rm D} \, \e^{-1/\lambda_{\alpha}} , 
     \label{eqTc} 
\end{equation} 
with $\lambda_{\alpha} = g \rho_{\alpha}(\mu)$.
We choose a value of the cutoff energy of bosons which mediate the pairing 
interactions as $\omega_{\rm D} = 1000 \, {\rm K}$. 
The value of $g$ is chosen so that $\Tc$ takes reasonable values 
such as $\sim 40 \, {\rm K}$ or $\sim 90 \, {\rm K}$ at the peak.

\begin{figure}[htb]
  \begin{center}
  \leavevmode \epsfxsize=7cm  \epsfbox{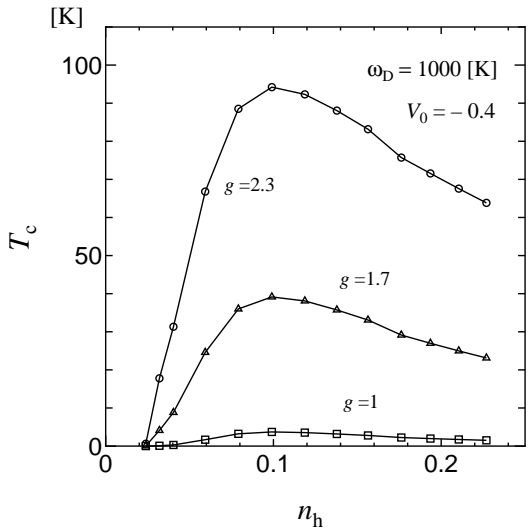}
  \end{center}
\caption{
Doping dependence of $\Tc$ for the $s$-wave pairing. 
Circles, triangles and squares indicate the results for $g = 2.3, 1.7$ 
and $1$, respectively. Solid lines are guides for the eyes. 
}
\label{fig:Tcnh_swave}
\end{figure}

\begin{figure}[htb]
  \begin{center}
  \leavevmode \epsfxsize=7cm  \epsfbox{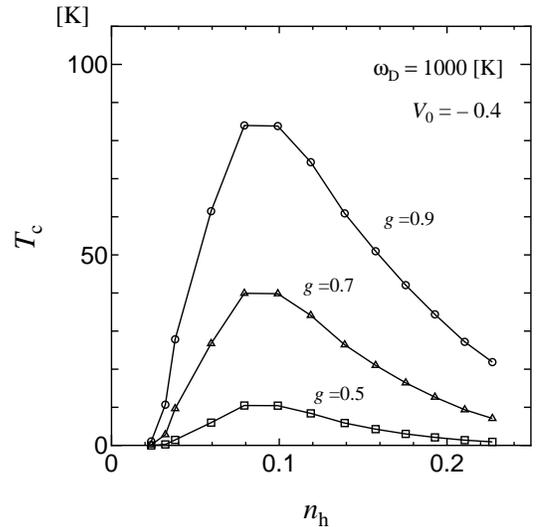}
  \end{center}
\caption{
Doping dependence of $\Tc$ for the $d$-wave pairing. 
Circles, triangles and squares indicate the results for $g = 0.9, 0.7$ 
and $0.5$, respectively. Solid lines are guides for the eyes. 
}
\label{fig:Tcnh_dwave}
\end{figure}

Figures~\ref{fig:Tcnh_swave} and \ref{fig:Tcnh_dwave} show the doping 
dependence of the $\Tc$ for the $s$-wave pairing and $d$-wave pairing, 
respectively. 
{From} the sensitivity of the singular exponential form of eq.~(\ref{eqTc}), 
we obtain marked suppressions of $\Tc$ near the AF boundary 
($\nh \gsim \nhc$) and thus peak structures around $\nh \sim 0.1 = 10 \%$. 
In particular, for the $d$-wave pairing, the peak is narrow, and seems to 
coincide with the experimental phase diagram of HTSC.

In order to be more precise, we should take into account the temperature 
dependence of the DOS in the estimation of $\Tc$. 
However, for $\nh \gsim \nhc$, the AF correlation length $\xi$ does not 
strongly depend on temperature at low temperatures~\cite{Bir89}. 
When its temperature dependence is ignored, 
the DOS are almost independent of the temperature. 
For example, it can be confirmed that the results for $T=0.005$ are 
almost the same as those for $T=0.01$. 
On the other hand, even when the temperature dependence of $\xi$ 
is taken into account, it does not change the qualitative result. 
It only emphasizes the peak of the $\Tc$, 
because the pseudogap becomes deeper for longer $\xi$ and it reduces 
$\Tc$ more.

\begin{figure}[htb]
  \begin{center}
  \leavevmode \epsfxsize=7cm  \epsfbox{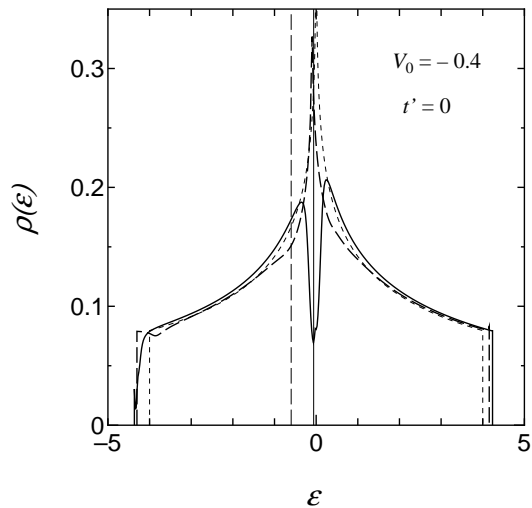}
  \end{center}
\caption{
DOS of the underdoped region 
$\mu = -0.06$ and $\nhc \approx 0.0237$ (solid line), 
and that of the overdoped region $\mu = -0.6$ and $\nhc \approx -0.227$ 
(broken line). 
The dotted line indicates the DOS when $V_0 = 0$. 
The vertical thin solid and broken lines indicate the chemical 
potentials $\mu = -0.06$ and $\mu = -0.6$, respectively. 
}
\label{fig:pseudogap}
\end{figure}

Figure~\ref{fig:pseudogap} shows the DOS of the underdoped region 
($\nh \approx 0.0237$) and that of the overdoped region 
($\nh \approx 0.227$). 
It is seen that the pseudogap is deep and clear in the underdoped region, 
while it becomes very shallow in the overdoped region. 
We also find that the pseudogap becomes large in the direction of 
$(\pm \pi,0)$ and $(0, \pm \pi)$, 
while it becomes small in the direction of 
$(\pm \pi, \pm \pi)$~\cite{ShiUP}.

It is straightforward to extend the present calculation to 
the $t' \ne 0$ case. 
When we assume $t' = -0.2$ and a similar doping dependence of the AF 
correlation length which decreases with doping, 
we also have a peak structure of the $\Tc$ near 
$\nh \approx 0.13$~\cite{ShiUP}. 
This peak structure is also consistent with the experimental phase 
diagrams except that the peak becomes very steep due to 
the van Hove singularity. 
However, it is easily verified by a phenomenological consideration 
that such steepness of the peak is not at all essential. 
If we assume that a large $t'$ is appropriate for the HTSC, 
the Fermi surface crosses the van Hove singularities 
at a large hole concentration. 
Thus, the DOS at the Fermi energy and $\Tc$ have a sharp peak around 
the hole concentration. 
However, such singular behavior or marked enhancement of the DOS 
with hole doping has not been observed in any experiments, 
for example, the specific heat and susceptibility measurements. 
This suggests that the singularity is eliminated finally by some 
many-body effect. 
If we take this many-body effect into our phenomenological model
beforehand, models with $t' \approx 0$ might be reasonable in order to
describe a realistic situation,
because it is consistent with the experimental facts that no singularity
occurs in the overdoped region, and the AF correlation length
increases near the half-filling.

We omit vertex corrections and dynamical effects. 
However, since these effects should reduce $\Tc$ in the underdoped 
region, it is likely that they only emphasize the peak structure of $\Tc$. 
Also, the strong coupling effects would not be essential regarding 
the qualitative behavior of the $\Tc$. 
The essential point is that $\Tc$ is very sensitive to the DOS, 
as explicitly written in the singular exponential form eq.~(\ref{eqTc}) 
within the weak coupling theory. 
The quantitative improvement of the theory by including these effects 
remains for future study.

In conclusion, we propose a mechanism which explains the peak structure 
in the doping dependence of $\Tc$ of HTSCs. 
It is shown that the pseudogap due to AF fluctuations suppresses $\Tc$ 
in the underdoped region and eventually destroys the superconductivity 
at a finite doping. 
On the other hand, the pseudogap phenomena is less pronounced 
near the optimum doping at which $\Tc$ is maximum. 
Then, DOS is large since the Fermi energy is at the shoulder of 
the van Hove singularity. This leads to high $\Tc$. 
The decrease of $\Tc$ in the overdoped region is due to the decrease of 
the DOS~\cite{Fuk87,Shi87,Koh99}, 
since the Fermi energy is distant from the van Hove singularity.

The authors would like to thank Professor D. Rainer and Professor 
J. Friedel for useful discussions and encouragements. 
This work was partially supported by a grant for CREST from JST.


\end{document}